# Diffusional Nucleation of Nanocrystals and Their Self-Assembly into Uniform Colloids

**Vladimir Privman**

Department of Physics, and Center for Advanced Materials
Processing, Clarkson University, Potsdam, NY 13699, USA

We review theoretical explanation of mechanisms of control of uniformity in growth of nanosize particles and colloids. The nanoparticles are synthesized as nanocrystals, by burst nucleation from solution. The colloids are self-assembled by aggregation of these nanocrystals. The two kinetic processes are coupled, and both are driven by diffusional transport. The interrelation of the two processes allows for formation of narrow-size-distribution colloid dispersions which are of importance in many applications. We review a mathematical model of cluster growth by capture of diffusing "singlets." Burst nucleation of nanoparticles in solution is then analyzed. Finally, we couple it to the secondary process of aggregation of nanoparticles to form colloids. We address aspects of modeling of particle size distribution, as well as other properties.

## 1. Introduction

In colloid and nanoparticle science, it is important to devise controlled synthesis approaches for obtaining uniform particles in solutions. The mechanisms can be actually different for colloids — suspensions of micron and sub-micron size particles, as compared to nanoparticles — those of sizes 0.01 µm (10 nm), and smaller. A broader goal of a theoretical modeling program includes understanding the kinetics of nucleation, growth, aggregation, and surface interactions of fine particles. Here we review modeling [1-9] of the process of burst-nucleation and diffusional growth of typically crystalline nanoparticles in solution, as well as of the accompanying secondary process of diffusional aggregation of these nanoparticles to form uniform polycrystalline colloids.

Uniform particle formation in solution, is an active field with many open problems and experimental as well as theoretical challenges. We have developed quantitative modeling [1,3,5-8] of the narrow size distributions observed for properly selected experimental conditions in synthesis of "monodispersed" (uniform) colloidal particles of various compositions. We have also addressed quantitatively [9] the nanoparticle size distribution in the model of burst nucleation, which, however, in its "classical" form is expected to be at best only approximately valid for real nanoparticle synthesis.

In Section 2, we generally address the particle size selection mechanism. In Section 3, we outline a mathematical treatment of diffusional growth by capture of monomers. Our model for burst nucleation of nanoparticles is presented in Section 4, in which we also survey the limitations of the model. When burst nucleation is accompanied by the secondary process of nanoparticle aggregation, self-assembly of uniform particles of colloid dimensions results. This two-stage process is surveyed in Section 5. Finally, in Section 6 we discuss additional developments and open problems, specifically the shape selection and shape distribution in fine particle synthesis.

## 2. Size selection in uniform particle synthesis

The concept of "monodispersed" colloid particles for applications, usually implies particle diameter distributions of relative width 6-12%. For nanosize particles, what do we mean by "monodispersed" at the nanoscale? It is expected that for nanotechnology applications, uniform size (and shape) really means "atomically identical." This is particularly true for future electronic devices. For many other applications, requirements for nanoparticle uniformity will also be strict.

Therefore, methods of controlling size and shape distributions, important for most applications of colloid suspensions, will be even more important for nanotechnology. Here we consider situations with "building blocks" from which particle are formed, as well as particles themselves, transported by diffusion in solution. The singlet (monomer) building blocks in nanoparticle synthesis in solution are atomic-size solute species (atoms, ions, molecules), whereas for colloid synthesis they are the (nanosize, typically nanocrystalline) primary particles. In the colloid case, the supply of singlets is "naturally" controlled by the parameters of their own burst nucleation. However, in principle the monomers for both processes can be also added/mixed in externally.

A particle size distribution of interest is illustrated in Figure 1. Mechanisms such as cluster-cluster aggregation or cluster ripening due to exchange of monomers, while making the size distribution grow, also broaden it: They cannot lead to narrow size selection. Indeed, most growth/coarsening mechanisms that involve diffusional transport broaden the distribution because larger particles have larger collection area for capturing "building blocks," as well as, e.g., for spherical particles, less surface curvature, which implies generally slightly better binding of monomers, resulting in less detachment.

Narrow particle size distribution can be achieved by several techniques. The simplest is to actually block the growth of the "right side" of the peak, cf. Figure 1, by "caging" the particles. An example could be nanoparticles grown inside nanoporous structures or objects. We do not consider this technique, which has been reviewed, e.g., in [10], and has a disadvantage of requiring the use of additional chemicals that later remain part of the formed particles.

Another approach involves dynamical processes that erode the left side of the peak, fast enough as compared to the peak broadening by coarsening processes, to maintain narrow distribution. The burst-nucleation process analyzed in Section 3, falls in this category. Unfortunately, other coarsening processes can eventually broaden the distribution after the initial nucleation burst. We will return to these issues in Section 3.

An important mechanism [1] that yields particle size distributions narrow on a relative scale, involves fast supply of monomers, of concentration $C(t)$, see Figure 1. The monomers "feed" the peak, thus pushing it to larger sizes, and the process can be fast enough not to significantly broaden the distribution on a relative scale, and, with proper time-dependent $C(t)$, not to generate too large a "shoulder" at small clusters. It is therefore quite natural to focus on the time dependence of the singlet (monomer) availability, and its impact on the size distribution of the products. Specifically, for nanosize particle preparation, there has been interest in stepwise processes, e.g., [11,12]: After achieving the initial nanoparticle distribution, batches of singlets are added to induce further growth.

Let $N_s(t)$ denote the density of particles consisting of $s$ singlets, at time $t$. We are interested in the situation illustrated in Figure 1: The distribution evolves in time with a peak eventually present at some relatively large $s$ values. Let us denote the singlet concentration by

$$C(t) \equiv N_1(t). \quad (1)$$

The singlets can be supplied as a batch, several batches, or at the rate $\rho(t)$, per unit volume. They are consumed by the processes involving the production of small clusters, in the "shoulder" in Figure 1. They are also consumed by the growing large clusters in the peak.

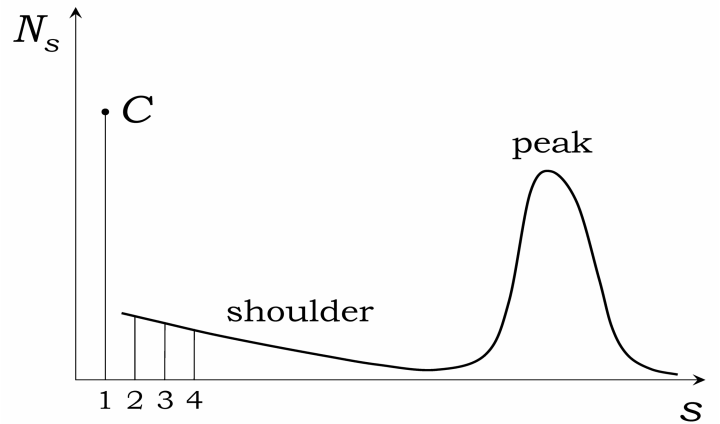

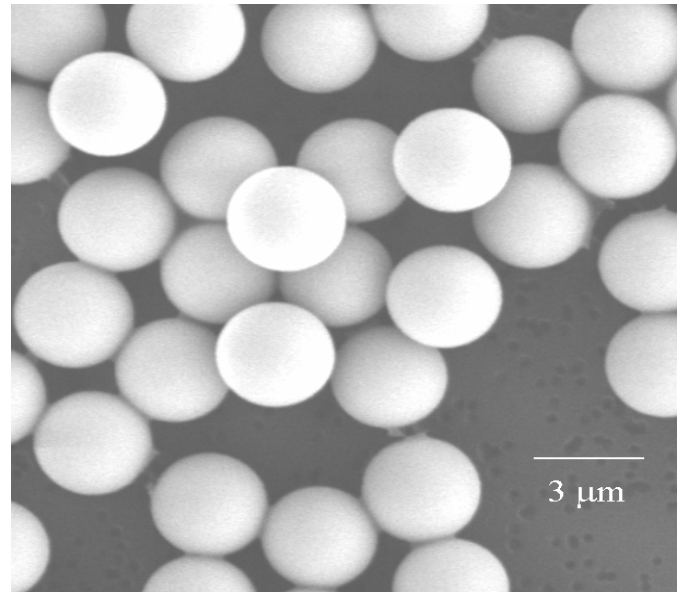

**Figure 1.** The top panel illustrates the desired particle size distribution. The peak at the larger cluster sizes can grow fast, at the expense of the singlets, which can be supplied externally. The distribution for $s > 1$ can be usually assumed a smooth function of $s$, though the vertical bars at $s = 1, 2, 3, 4$ emphasize that the $s$ values are actually discrete. The bottom panel: SEM image of polycrystalline spherical CdS colloid particles illustrating the attainable uniformity of the size and shape distribution.

There are two issues to consider: How is the peak created in the first place, and how to grow it without much broadening. In the next section, we address some mathematical aspects of the latter issue. Regarding the former issue, for nanoparticle synthesis the main mechanism of the early formation of the peak is by burst nucleation, when nuclei of sizes larger than the critical size form by growing over the nucleation barrier. Of course, *seeding* is another way of initiating the peaked size distribution both for colloid and nanoparticle growth. For colloid synthesis without seeding, the initial peak formation is more subtle and could actually be a result of few-singlet cluster-cluster aggregation at the early growth stages, as further mentioned at the end of Section 6.

## 3. Singlet-driven particle growth

We consider a mathematical model of growth dominated by irreversible capture of singlets by the larger growing aggregates. We use the rate equations, with $\Gamma_s$ denoting the rate constants for singlet capture by the $s \geq 1$ aggregates,

$$\frac{dN_s}{dt} = (\Gamma_{s-1} N_{s-1} - \Gamma_s N_s)C, \quad s > 2, \quad (2)$$

$$\frac{dN_2}{dt} = (\frac{1}{2}\Gamma_1 C - \Gamma_2 N_2)C, \quad (3)$$

$$\frac{dC}{dt} = \rho - \sum_{s=2}^{\infty} s \frac{dN_s}{dt} = \rho - \Gamma_1 C^2 - C \sum_{s=2}^{\infty} \Gamma_s N_s. \quad (4)$$

The approximation that the only process involving the $s > 1$ aggregates is that of capturing singlets at the rate proportional to the concentration of the latter, $\Gamma_s C$, has been commonly used, e.g., [1,5-6,13-15]. We will comment on elaborations later. More complex processes, such as cluster-cluster aggregation [16,17], detachment [2,4] and exchange of singlets (ripening), etc., also contribute to particle growth. However, in colloid synthesis they are much slower than the singlet-consumption growth. In addition, they broaden the particle size distribution.

Another approximation involved in writing Eqs. (2-4) is that of ignoring particle shape and morphology distribution. We avoid this issue, which is not well understood, by assuming that the growing aggregates rapidly restructure into compact bulk-like particles, of an approximately fixed shape, typically, but not always, spherical for colloids. This has been experimentally observed in uniform colloid synthesis [1,18-23]. Without such restructuring, the aggregates would be fractal [17,24]. in Section 6, we comment on the shape selection issue as an unsolved problem.

For nanosize particle formation, the assumptions in the present approximation that should be scrutinized are those of ignoring singlet detachment, and "embryo" breakup, for the particles in the shoulder in Figure 1. Indeed, unlike colloid growth, which is fast and irreversible for all $s$ in solution synthesis processes, the nanosize particle growth will be typically held back by a nucleation barrier [1,6,9,11]. During the late stage growth, that follows the initial nucleation burst [9,25,26], the barrier can be quite high. The distribution in the shoulder will approach the equilibrium Boltzmann form, governed by the excess free energy of the aggregate formation. It is interesting to note that this fast equilibration means that the singlets "stored" in the small, "shoulder" aggregates will be ultimately available for consumption by larger aggregates in the peak. Burst nucleation is analyzed in Section 4.

Here we focus on the situations for which the assumptions leading to Eqs. (2-4) apply: "Minimal" models of colloid growth and certain stepwise nanoparticle growth processes. If the singlets are supplied/available constantly, then the distribution, both for colloids and nanoparticles, will develop a large shoulder at small aggregates, with no pronounced peak at $s \gg 1$. If the supply is limited, then only small aggregates will be formed. Our key recent discovery in studies of colloid synthesis [1,6] has been that there exist protocols of singlet availability, at the rate $\rho(t)$ which is a slowly decaying function of time, that yield peaked (at large sizes) distributions at large times. Furthermore, the primary (nanocrystal nucleation) process in uniform polycrystalline colloid synthesis, naturally "feeds" the secondary process (of nanoparticles aggregation to form colloids) just at a rate like this.

Solution of Eqs. (2-4) requires numerical approaches and is not particularly illuminating as to the nature of the particle growth. Therefore, to explore the nature of the peak growth, in this section we will introduce several additional assumptions which will allow us to go a long way in simplifying the problem in closed analytical form. The main idea is that, once the peak is formed after some transient time or by seeding, the particles in the peak are the main consumers of the available singlets.

This assumes that the singlet concentration is controlled by adding them externally [6,11,12]. For nanoparticles, the addition should be at such a rate that the nucleation barrier remains high. The shoulder will then adjust to assume an approximately equilibrium shape, but the production of new larger, supercritical aggregates will be negligible. For colloid growth, the shoulder will also evolve, with new particles generated. However, if the number of larger aggregates is already significant, they will dominate the consumption of singlets.

In order to understand how a well-developed peak can evolve while remaining relatively narrow, let us entirely inhibit generation of new small aggregates, by setting

$$\Gamma_1 \to 0, \tag{5}$$

which is an approximation appropriate for the later-time regime when a well-developed peak already exists and particles in it are the main "consumers" of singlets, whereas production of new small particles into the shoulder of the distribution, see Figure 1, which also occurs by consumption of singlets, no longer plays any significant role. Furthermore, we will assume that $s$ is a continuous variable, since we are interested here in $s \gg 1$, and that it varies in the range $0 \le s < \infty$.

For calculations assuming singlet transport by diffusion, one can take the large-$s$ Smoluchowski expression for the rates [2,27-28],

$$\Gamma_{s \gg 1} = \Upsilon s^{1/3}, \tag{6}$$

where $\Upsilon$ is a known constant. Note that $\Gamma_{s \gg 1}$ is proportional to the aggregate linear dimension (which yields the factor $s^{1/3}$) times the singlet diffusion constant. The results in this section actually apply for general $\Gamma_s$.

The last approximation is introduced while deriving the continuous-$s$ form of Eq. (2): We retain only the leading $s$ derivative, ignoring the "diffusive" second-derivative term (this will be revisited later, in Section 4). The consequences of this approximation, used, e.g., in [6,13], will be discussed later. Thus, we replace Eq. (2) by

$$\frac{\partial N(s,t)}{\partial t} = -C(t)\frac{\partial}{\partial s}[\Gamma(s)N(s,t)], \tag{7}$$

with Eq. (4) replaced by

$$\frac{dC(t)}{dt} = \rho(t) - C(t)\int_0^\infty ds[\Gamma(s)N(s,t)]. \tag{8}$$

Let us define

$$\tau(t) = \int_0^t dt' C(t') \ge 0, \tag{9}$$

and introduce the function $u(s,\tau)$ via the relation

$$\tau = \int_u^s \frac{ds'}{\Gamma(s')}. \tag{10}$$

We point out that usually $\Gamma(s') > 0$, and the lower limit of integration can be taken to zero. The asymptotic rate in Eq. (6) does vanish at argument 0, because of our cavalier treatment of the small-$s$ behavior. However, the integral happens to converge, so no additional care is needed. We can safely define the quantity $s_{\min}(\tau)$ via

$$\tau = \int_0^{s_{\min}} \frac{ds'}{\Gamma(s')}. \tag{11}$$

As $u$ is increased from zero to infinity, the corresponding $s(u,\tau)$, for fixed $\tau$, increases from $s_{\min}(\tau)$ to infinity.

Next, we notice that the relation between the differentials implied by Eq. (10), namely,

$$d\tau = \frac{ds}{\Gamma(s)} - \frac{du}{\Gamma(u)}, \tag{12}$$

allows us to calculate partial derivatives in terms of $\Gamma(s)$ and $\Gamma(u) = \Gamma(u(s,\tau(t)))$. This, in turn, allows one to verify, by a cumbersome calculation not reproduced here, that Eq. (7) is solved by

$$N(s,t) = \frac{\Gamma(u(s,\tau(t)))}{\Gamma(s)} N(u(s,\tau(t)),0),$$
$$\text{for } s \ge s_{\min}(\tau(t)), \tag{13}$$

and

$$N(s,t) = 0, \quad \text{for } 0 \le s \le s_{\min}(\tau(t)), \tag{14}$$

where the discontinuity at $s_{\min}(\tau(t))$ is possible if the initial distribution at time zero, $N(s,0)$, is nonzero at $s = 0$. Actually, within the present approximation of ignoring the effects of the details of the size distribution for small $s$, we could as well set $N(0,0) = 0$.

Let us summarize the above observations by emphasizing that we consider a particle size distribution which at time $t = 0$ already has a well-developed significant peak at large cluster sizes. Equations (13-14) will provide an approximate description of further evolution of this peak with time, due to supply of singlets at the rate $\rho(t)$. The form of the distribution at small particle sizes plays no role in the derivation. In fact, neglecting the second-derivative in $s$, "diffusive" term in writing Eq. (7), leads to certain artificial features. Specifically, sharp corners and discontinuities of the initial distribution (as well as its derivatives, etc.) will not be smoothed out. The fact that the initial distribution is only meaningful for $s \geq 0$ translates into the sharp cutoff at $s_{\min}$ for times $t > 0$. Had we included the diffusive term, the distribution would extend smoothly to $s = 0$ for all times. However, no closed-form analytical solution would be available. While this lack of smoothness is probably not important for a semi-quantitative evaluation of the size distribution, one aspect should be emphasized as critical: If the initial distribution is already very sharp, then the neglect of the diffusive term in our expressions may result in underestimating the width of the evolving peak.

To complete the description of the particle size distribution within the non-diffusive approximation, we have to discuss the estimation of the function $\tau(t)$. Equations (8-9) can be rewritten, using Eq. (13), as a system of coupled differential equations for two unknown functions $\tau(t)$ and $C(t)$, with $\tau(0) = 0$, and $C(0)$ externally controlled,

$$\frac{d\tau}{dt} = C(t), \tag{15}$$

$$\frac{dC}{dt} = \rho(t) - C(t) F(\tau), \tag{16}$$

where

$$F(\tau) = \int_{s_{\min}(\tau)}^{\infty} ds \left[ \Gamma(u(s,\tau)) N(u(s,\tau),0) \right]. \tag{17}$$

These equations are easily programmed for numerical evaluation, especially if the function $F(\tau)$ is calculable analytically, so that numerical integration can be avoided. The latter might be possible for the power-law rate in Eq. (6), provided the initial distribution $N(s,0)$ is not too complicated.

Within the approximation developed here, the number of particles larger than singlet, $M$, obviously remains constant,

$$M = \int_{s_{\min}(t)}^{\infty} ds N(s,t) = \int_0^{\infty} ds N(s,0). \tag{18}$$

The change in the average size of the particles larger than singlet,

$$\langle s \rangle_t = \frac{1}{M} \int_{s_{\min}(t)}^{\infty} ds [s N(s,t)], \tag{19}$$

can be evaluated directly from $C(t)$,

$$\langle s \rangle_t = \langle s \rangle_0 + \frac{1}{M}[C(0) - C(t) + \int_0^t dt' \rho(t')]. \tag{20}$$

Furthermore, consideration of the increment relations following from Eq. (12), suggests that the growth of the width of the peak, $W_t$, can be roughly estimated from

$$W_t \approx \frac{\Gamma(\langle s \rangle_t)}{\Gamma(u(\langle s \rangle_t, \tau(t)))} W_0 > W_0, \tag{21}$$

where the inequality follows from Eq. (10), assuming that for large $s$, $\Gamma(s) > 0$ is an increasing function. This excludes an important case of constant $\Gamma$, appropriate for certain models of polymerization. In that case, however, Eqs. (2-4) can be analyzed directly [14,15], so that the present formulation is not needed.

In connection with Eq. (21), we note that additional broadening will result from the second-derivative "diffusive" term neglected in our continuous-$s$ equations. The model with the diffusive term included, requires serious numerical efforts, as does the original, discrete-$s$ model; however, see Section 4 for some explicit expressions.

In summary, with the reservations regarding the width (under)estimates, numerical calculation of the functions $\tau(t)$ and $C(t)$, via Eqs. (15-17), goes a long way in estimating various properties of the growing, peaked size distribution. Even at the level of the approximations leading to Eq. (21), it is obvious that the size distribution never actually narrows in absolute terms. Specifically, experimentally realized monodispersed particle synthesis procedures in solution, in the colloid domain, actually yield small relative peak width, $W_t / \langle s \rangle_t$, by utilizing fast increase in $\langle s \rangle_t$ via consumption of singlets, on the time scales too short for the "diffusive" broadening to set in.

### 4. Burst nucleation

The model of burst nucleation [9,25,26] is appropriate for nanosize particles, typically, crystals, consisting of $n$ monomers (we will reserve $s$ for the count of singlets in growth of colloids, considered in Sections 5-6). The larger particles, with $n > n_c$, where $n_c$ is the critical cluster size, irreversibly capture atom, ion or molecule singlets which are diffusing solutes. However, the dynamics in the shoulder, for $n < n_c$, see Figure 2, vs. Figure 1, is no longer ignored: The subcritical ($n < n_c$) aggregates are assumed instantaneously rethermalized.

Thus, it is assumed that in a supersaturated solution with time-dependent monomer concentration $c(t)$, thermal fluctuations cause formation of aggregates (embryos), controlled by the free-energy barrier imposed by the surface free energy. The full dynamics of these few-atom clusters involves complicated transitions between embryos of various sizes, shapes, as well as internal restructuring. These processes are presently not well understood. However, the dynamics of embryos is fast, and their sizes are approximately thermally distributed and modeled by a Gibbs-like form [1,5] of the free energy of an $n$-monomer embryo,

$$\Delta G(n,c) = -(n-1)kT \ln(c/c_0) + 4\pi a^2 \left(n^{2/3} - 1\right)\sigma, \tag{22}$$

where $k$ is Boltzmann constant, $T$ is the temperature, $c_0$ is the equilibrium concentration of monomers, and $\sigma$ is the effective surface tension.

The first term is the bulk contribution. It is derived from the entropy of mixing of noninteracting solutes and is negative for $c > c_0$, therefore favoring larger clusters. The second, positive term represents the surface free-energy, proportional to the area, $\sim n^{2/3}$. The effective solute radius, $a$, is defined in such a way that the radius of an $n$-solute embryo is $an^{1/3}$. It can be estimated by requiring that $4\pi a^3/3$ equal the unit-cell volume per singlet (including the surrounding void volume) in the bulk material.

As in most treatments of homogeneous nucleation, we assume that the distribution of aggregate shapes can be neglected: A "representative" aggregate is assumed spherical in the calculation of its surface area and the monomer transport rate to it. We note that even the surface tension of spherical particles varies with their size. This effect, as well as any geometrical factors that might be needed because real clusters are not precisely spherical, is neglected. The effective surface tension of nanoparticles is only partially understood at present [29]. Thus, $\sigma$ can be either assumed [1,5,7,8] close to $\sigma_{\text{bulk}}$, or fitted as an adjustable parameter.

The free energy, Eq. (22), increases with $n$ until it reaches the "peak of the nucleation barrier" at $n_c$,

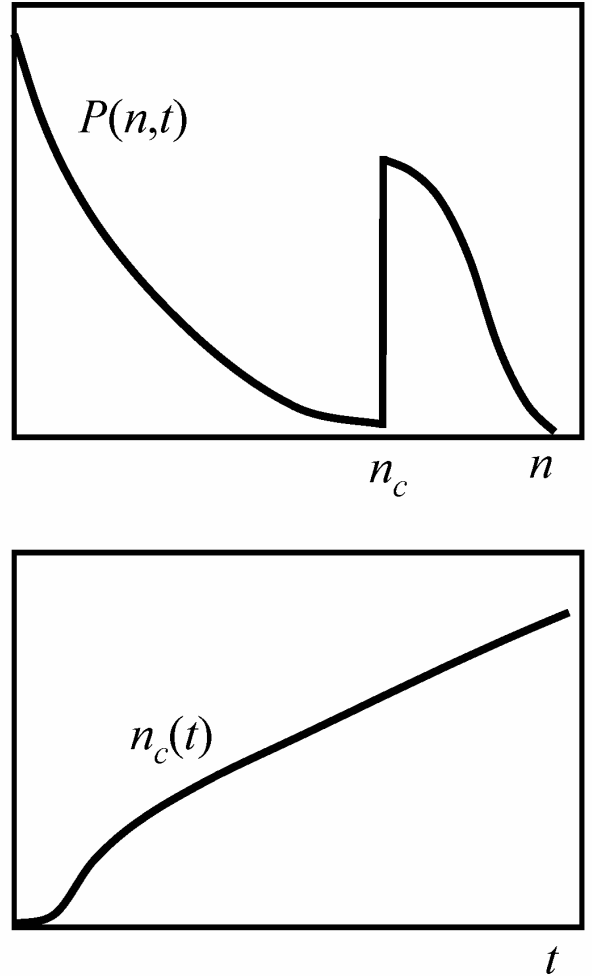

**Figure 2.** The top panel schematically illustrates the large-time form of cluster size distribution in burst nucleation. The bottom panel sketches the time dependence of the critical cluster size, showing the induction period, followed by the "burst," and then the asymptotically linear growth.

$$n_c(c) = \left[\frac{8\pi a^2 \sigma}{3kT \ln(c/c_0)}\right]^3. \tag{23}$$

For $n > n_c$, the free energy decreases with $n$. However, the kinetics then becomes irreversible and is no longer controlled by $\Delta G$.

The specific property of burst nucleation is that the barrier, and $n_c$, strongly depend on the monomer concentration, $c$, which leads to a significant suppression of nucleation after the initial burst, during which $c/c_0$ decreases from its initial value $c(0)/c_0 \gg 1$ to it asymptotic large-time equilibrium value 1. The large-time form of the particle size distribution in burst nucleation is shown in Figure 2. Specifically, the embryonic matter below $n_c$ is thermalized on time scales much faster than those of other dynamical processes, so that the concentration of embryos, with sizes in $dn$, is given by $P(n,t)dn$, with the particle size distribution

$$P(n < n_c, t) = c(t) \exp\left[\frac{-\Delta G(n, c(t))}{kT}\right], \quad (24)$$

where $n_c = n_c(c(t))$.

The rate of production of supercritical clusters, to be denoted by $\rho(t)$ for use in Section 5, is then expressed [1] as

$$\rho(t) = K_{n_c} c P(n_c, t) = K_{n_c} c^2 \exp\left[\frac{-\Delta G(n_c, c)}{kT}\right], \quad (25)$$

where $K_n = 4\pi a n^{1/3} D$ is the Smoluchowski expression [2,27,28] for the rate of irreversible intake of diffusing solutes by growing spherical clusters. We already encountered this rate in Eq. (6). Here we use the large-$n$ form for supercritical clusters, $n \geq n_c \gg 1$, and $D$ is the diffusion coefficient for monomers in a solution with viscosity $\eta$. $D$ can be estimated as $\sim kT/6\pi\eta a$, up to geometrical factors (the effective unit-cell-derived radius $a$ must be replaced by the hydrodynamic radius for diffusing monomers).

Although real clusters undergo both attachment and detachment of monomers (with detachment still present at sizes above $n_c$), we model the expected rapid growth of the supercritical, $n > n_c$, clusters within the approximation of irreversible capture of diffusing monomers,

$$\frac{\partial P(n,t)}{\partial t} = (c(t) - c_0)(K_{n-1} P(n-1, t) - K_n P(n, t)). \quad (26)$$

Comparing to Eq. (2), the difference $c(t) - c_0$ is used here in place of $c(t)$ to ensure that the growth of clusters stops when the equilibrium concentration $c_0$ is reached.

The variation of the nanocluster surface tension with its radius, mentioned above, is accompanied by a variation of the effective equilibrium concentration, $c_0$, with radius, which gives rise to Ostwald ripening [30]. This, as well as other possible coarsening processes, such as cluster-cluster aggregation [16,17], are neglected here because burst nucleation is expected [1,9] to be a much faster process. However, for large times such coarsening processes will gradually widen the particle distributions seen in experiment and slow down the growth of the particle size, which, as will be argued shortly, for burst nucleation alone is well characterized by the function $n_c(t)$ schematically shown in Figure 2. Furthermore, in some situations the large-time asymptotic linear behavior has a very small slope [31], so de-facto the growth would "freeze" if it were due to burst-nucleation alone.

We further comment that in addition to growth (shrinkage) by attachment (detachment) of monomers, clusters of all sizes can undergo internal restructuring, a complex phenomenon the modeling of which for nanoscale clusters is only in its early stages [32,33]. Without such restructuring, the clusters would grow according to diffusion-limited aggregation or similar processes and could be fractals [16,17], whereas observations of the density and X-ray diffraction data of colloidal particles aggregated from burst-nucleated nanocrystalline subunits indicate that their polycrystalline structure has the density of the bulk [1,34]. There is primarily experimental, but also modeling evidence [1,4,5,7,8], that for larger clusters such restructuring leads to compact particles with smooth surfaces, which then grow largely irreversibly.

The "right side" of the supercritical distribution, see Figure 2, grows towards larger clusters by capturing monomers, but, at the same time, its "left side" is eroded by the thermalized subcritical distribution which extends up to $n_c(t)$ — a monotonically increasing function of time. The form of the supercritical distribution depends on the initial conditions. As will be demonstrated shortly, at large times it will eventually have its maximum at $n = n_c$, and will take on the form of a truncated Gaussian. This is illustrated in Figure 2, where the peak of the full Gaussian curve (not shown) is actually to the left of $n_c$.

Numerical results for time-dependent distributions and for several initial conditions, presented in [9], were obtained by a novel efficient numerical integration scheme which is not reviewed here. In what follows, we concentrate on the derivation of analytical results for large times. We note that one must be consistent, in both the asymptotic and numerical treatments, with the conventions for relating the discrete-$n$ quantities, such as the monomer concentration $c(t)$, to the values of the continuous distributions. We have chosen the simple convention $c(t) = P(1, t)$, rather than, e.g., a convention to treat the monomer concentration $c(t)$ separately of the rest of the distribution, as was done in Section 3. Then the conservation of matter is expressed by that the quantity

$$\int_1^{n_c} n\, c(t) \exp\left[\frac{-\Delta G(n,c(t))}{kT}\right] dn + \int_{n_c}^{\infty} n\, P(n,t)\, dn \qquad (27)$$

remains constant as a function of time.

It can be shown that for large times the kinetic equations suggest an asymptotic parameterization of the form

$$P_G(n,t) = \zeta(t) c_0 \exp\left[-(\alpha(t))^2 (n - K(t))^2\right], \qquad (28)$$

for $n > n_c(t)$ and large $t$. We also define the "peak offset"

$$L(t) \equiv n_c(t) - K(t). \qquad (29)$$

The asymptotic analysis starts with writing Eq. (26) in a continuous-$n$ form. Unlike Section 3, here we are interested in the precise peak shape and therefore we keep terms up to the second derivative,

$$\frac{\partial P}{\partial t} = (c - c_0)\left[\left(\frac{1}{2}\frac{\partial^2}{\partial n^2} - \frac{\partial}{\partial n}\right)(K_n P)\right]. \qquad (30)$$

This describes the irreversible growth of clusters above the critical size, where, within the assumption of the narrow Gaussian, $P(n,t)$ takes on appreciable values only over a narrow range. Thus we can approximate, for evaluation of the asymptotic behavior, $K_n \approx K_{n_c} = \kappa (n_c(t))^{1/3}/c_0$, where $\kappa \equiv 4\pi c_0 a D$. In terms of the dimensionless quantity

$$x(t) \equiv c(t)/c_0, \qquad (31)$$

we get

$$\frac{\partial P}{\partial t} = \kappa(x(t) - 1)(n_c(t))^{1/3}\left(\frac{1}{2}\frac{\partial^2}{\partial n^2} - \frac{\partial}{\partial n}\right) P. \qquad (32)$$

From Eq. (23), in the large-time limit, when $c(t) \to c_0$, we have $x(t) - 1 \propto (n_c(t))^{-1/3}$, which cancels the factor $(n_c(t))^{1/3}$ in Eq. (32). For later convenience we introduce the constant $z$ via

$$z^2 \equiv \frac{64\pi^2 a^3 \sigma c_0 D}{3kT}. \qquad (33)$$

With this definition, for large times Eq. (32) then reduces to

$$\frac{\partial P}{\partial t} = \frac{z^2}{2}\left(\frac{1}{2}\frac{\partial^2}{\partial n^2} - \frac{\partial}{\partial n}\right) P. \qquad (34)$$

Substituting the Gaussian Eq. (28) into Eq. (34), establishes that the solution is indeed of the conjectured form and yields [9] the following asymptotic results for the parameters:

$$\alpha(t) \simeq 1/\sqrt{z^2 t}, \quad K(t) \simeq z^2 t/2, \quad \zeta(t) \simeq \Omega/\sqrt{z^2 t}. \qquad (35)$$

The preafctor $\Omega$ cannot be determined from the asymptotic analysis alone, because the overall height of the distribution is obviously expected to depend on the initial conditions.

The asymptotic behavior of the peak offset, Eq. (29), follows from the conservation of matter. Indeed, for large times the second term in Eq. (27) will be approximated by

$$\int_{n_c(t)}^{\infty} n P_G(n,t)\, dn, \qquad (36)$$

which must approach a constant value, equal to the initial total matter less the matter that remains in the thermal distribution as $c \to c_0$. The rather complicated mathematical analysis that follows, will not be reproduced here; see [9]. The key result is that conservation of matter implies

$$L(t) \propto \sqrt{t \ln t} \qquad (37)$$

for large times. Therefore, the leading asymptotic behavior of the critical cluster size is the same as that for $K(t)$,

$$n_c(t) \simeq z^2 t/2. \qquad (38)$$

Since the width of the truncated Gaussian is still given by $1/\alpha \sim \sqrt{t}$, we note that our results suggest linear growth of the distribution for large times, see Figure 2, with the relative width decreasing with time as $\sim t^{-1/2}$. Finally, one can show [9] that the difference $c(t) - c_0$ approaches zero as $\sim t^{-1/3}$.

We comment that the Gaussian distribution has provided a good fit at intermediate and large times for numerical data for various initial conditions, including for initially seeded distributions; see [9]. Numerical simulations also confirm the other expected features of burst nucleation, summarized in Figure 2: The initial induction period followed by growth "burst" that precedes the onset of the asymptotically linear growth.

It is experimentally challenging in many situations to unambiguously quantify the size distribution of nucleated nanocrystals, because of their tendency to aggregate, their distribution of non-spherical shapes, and other factors. Still, it is commonly found (and expected) in experiment that the distribution is two-sided around the peak, and that the final particles stop growing after a certain time. Both of these experimental observations are at odds with the predictions of the burst-nucleation model, and the discrepancies can be attributed to the assumed instantaneous thermalization of the clusters below the critical size. At very small sizes, below a cutoff value, which can be speculated to correspond to $n_{th} \approx 15\text{-}25$ monomers [6-8,35-37] (atoms, molecules, sub-clusters), structures can evolve very rapidly, so that the assumption of fast, thermally driven restructuring is justified.

At larger sizes, however, embryos can be expected to undergo a transition in which their internal atoms assume a more stable, bulk-like crystal structure, and they no longer restructure as easily, except perhaps at their surface layers. Thus for times for which $n_c(t) > n_{th}$, the "classical" nucleation model should be regarded as approximate. Modifications of the model have been contemplated in several previous studies of nucleation [9,38,39]. This, however, requires introduction of new parameters which are not as well defined and as easily experimentally accessible as those of the "classical" nucleation model. In fact, one of the most interesting applications of our present theoretical developments would be to try to estimate, based on experimental data, the deviations from the "classical" behavior and thus obtain information on the value of $n_{th}$ — the nanostructure size beyond which a "bulk-material" core develops. A similar effect in colloid synthesis will be mentioned in Section 6.

The extent to which our (unmodified) model describes the initial burst, as well as the range of applicability of the prediction of linear growth of $n_c(t)$, are interesting topics to explore further. We recall that other processes at all cluster sizes, such as cluster-cluster aggregation and ripening, can also modify the kinetics of the distribution, albeit these are usually expected to play role at time scales much larger than the initial nucleation burst.

## 5. Synthesis of uniform colloids

As described in the preceding section, the burst-nucleation mechanism, which ideally can yield narrow size distributions, is never realized in practice for extended growth times. For larger particles, nucleated in the initial burst and then grown to dimensions typically over several tens of nanometers in diameter, other growth mechanisms usually broaden the size distribution. Here we consider the two-stage mechanism whereby the nanosized primary particles, burst-nucleated and growing in solution, themselves become the singlets and are "consumed" by the singlet-driven aggregation that results in uniform secondary particles of colloid dimensions. The primary process is of the type considered in Section 4, whereas the secondary process is the one introduced in Section 3.

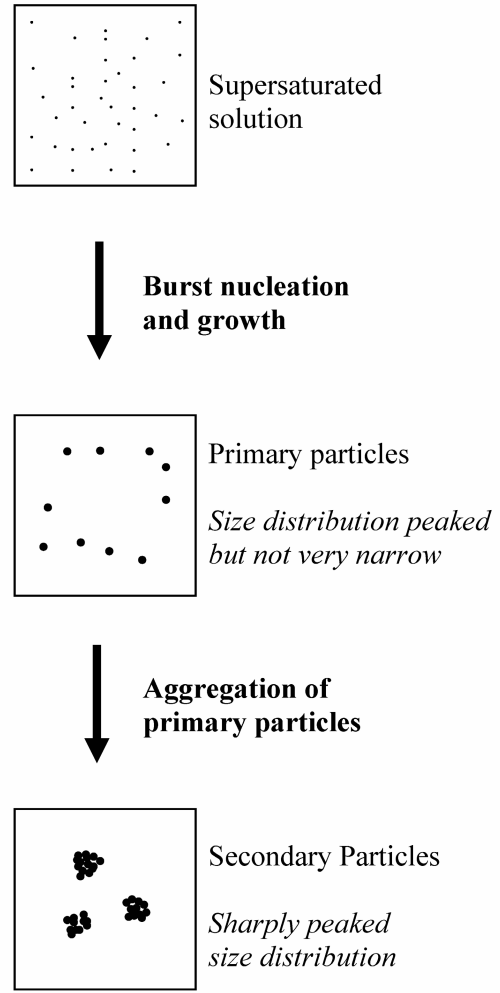

**Figure 3.** The two-stage synthesis mechanism of uniform colloids by self-assembly of diffusing aggregating nanocrystalline precursor subunits which are, in turn, formed by burst nucleation in a supersaturated solution, likely followed by additional growth/coarsening.

A large number of dispersions of uniform colloid particles of various chemical composition and shape, ranging in size from fraction of a micron to few microns, have been synthesized via the two-stage route [1,7-8,18-23,34,40-56]. Indeed, it has been found that many spherical particles precipitated from solution showed polycrystalline X-ray characteristics, such as ZnS [42], CdS [7,8,41], $Fe_2O_3$ [40], Au and other metals [1,23,52-54,56], etc. These particles are not single crystals. Rather, several experimental techniques have confirmed that most monodispersed colloids consist of small crystalline subunits [1,7-8,18-23,34,40-56]. Furthermore, experiments have observed [1,23,50] that the crystalline subunits in the final particles were of the same size as the diameter of the precursor singlets of sizes of order up to a couple of 10 nm, formed in solution, thus suggesting an aggregation-of-subunits mechanism. This two-stage growth process is summarized in Figure 3. The composite structure has also been identified in uniform non-spherical colloid particles [40,46-48,55], albeit perhaps thus far not as definitively as for the spherical case.

Here we review the simplest (in that it avoids introduction of unknown microscopic parameters) model that involves the coupled primary and secondary processes. Even this model requires numerical calculations and cannot be analyzed in closed form. In Section 6, we describe some improvements of the model that allow for better agreement with experimental observations. Additional details, examples of experimental parameters and results, and well as sample numerical data fits can be found in [1,5,7,8,57].

The reader might recall that in modeling the burst nucleation process in Section 4, the supercritical distribution was described by Eq. (26). To calculate the subcritical distribution, one only requires an expression for the time derivative $dc/dt$, since the whole subcritical distribution can be calculated if we know $c(t)$, see Eq. (24). We did not review, but only referenced our work [9] for mathematical steps, involving the conservation of matter, that give Eq. (37) and also yield a complicated expression for $dc/dt$ (not shown).

When the burst-nucleated supercritical particles are largely consumed by the secondary aggregation, we can instead assume that these primary particles are captured fast enough by the growing secondary particles so that the effect of their aging on the concentration of solutes can be ignored. Furthermore, the radius of the captured primary particles will be assumed close to the critical radius. We discuss the implications of these assumptions shortly. For now, we write our first expression that applies for the two-stage process, but does not apply to burst nucleation alone,

$$\frac{dc}{dt} = -n_c \rho. \qquad (39)$$

Recall that the rate of supercritical particle production, $\rho(t)$, was defined in Eq. (25) and is a known function of $c(t)$. Thus, Eq. (39), which expresses our approximation that the concentration of solutes is depleted solely due to the irreversible formation of the critical-size nuclei, yields

$$\frac{dc}{dt} = -\frac{2^{14}\pi^5 a^9 \sigma^4 D_a c^2}{(3kT)^4 [\ln(c/c_0)]^4} \exp\left\{-\frac{2^8 \pi^3 a^6 \sigma^3}{(3kT)^3 [\ln(c/c_0)]^2}\right\}, \qquad (40)$$

$$\rho(t) = \frac{2^5 \pi^2 a^3 \sigma D_a c^2}{3kT \ln(c/c_0)} \exp\left\{-\frac{2^8 \pi^3 a^6 \sigma^3}{(3kT)^3 [\ln(c/c_0)]^2}\right\}, \qquad (41)$$

which can be used to numerically calculate $\rho(t)$. The notation for various quantities here is the same as in Section 4. However, we denoted by $D_a$ the diffusion constant of the solutes, in order to distinguish it from that of the supercritical primary particles that constitute the "singlets" for the secondary process, to be denoted $D_p$.

The growth of the secondary (colloid) particles is facilitated by the appropriate chemical conditions in the system: The ionic strength and/or pH must be kept in ranges such that the surface potential approaches the isoelectric point, resulting in reduction of electrostatic barriers, thus promoting fast irreversible primary particle attachment. Formation of the secondary particles is clearly a diffusion-controlled process [1,18-23].

We describe the process by the equations for the distribution of growing particles by their size, cf. Eqs. (1-4). Here it is assumed that the particles are spherical, with the density close to that of the bulk material. Experimentally, the growing particles rapidly restructure to assume the final shape and density: They are not fractal even though the transport of the constituent units is diffusional. The modeling of this restructuring is an interesting unsolved problem on its own, but, as long as the restructuring is fast, its mechanism plays no role in formulating the model equations.

The cluster size $N_{s=1,2,3,...}(t)$ will be defined by how many primary particles (singlets) were aggregated into each secondary particle. The notation here is similar to that in Sections 2 and 3. For example, Eqs. (2-4) can be solved numerically with the initial conditions $N_{s=1,2,3,...}(0) = 0$. The simplest choice of the rate constants is the Smoluchowski expression

$$\Gamma_s \approx 4\pi R_p D_p s^{1/3}, \qquad (42)$$

where $R_p$ is the primary particle radius, and the approximate sign is used because several possible improvement to the simplest formula can be offered, as will be described below. A typical numerical calculation for a model of this type is shown in Figure 4, illustrating the key feature — size selection — the "freezing" of the growth even for exponentially increasing times (here is steps ×10).

Let us now discuss some of the numerous simplifying assumptions made in the model just formulated. We will also consider possible modifications of the model. In fact, Figure 4, which was based on one of the sets of the parameter values used for modeling formation of uniform spherical Au particles, already includes some of the modification; see [5] for details.

We note that since the assumption $s \gg 1$ is not applicable, the full Smoluchowski rate expression [2,27,28] should be used, which, for aggregation of particles of sizes $s_1$ and $s_2$, on encounters due to their diffusional motion, is

$$\Gamma_{s_1, s_2 \to s_1 + s_2} \simeq 4\pi \left[ R_p \left( s_1^{1/3} + s_2^{1/3} \right) \right] \left[ D_p \left( s_1^{-1/3} + s_2^{-1/3} \right) \right], \quad (43)$$

where for singlet capture $s_1 = s$ and $s_2 = 1$. This relation can not only introduce nontrivial factors for small particle sizes, as compared to Eq. (42), but it also contains an assumption that the diffusion constant of $s$-singlet, dense particles is inversely proportional to the radius, i.e., to $s^{-1/3}$, which might not be accurate for very small, few-singlet aggregates.

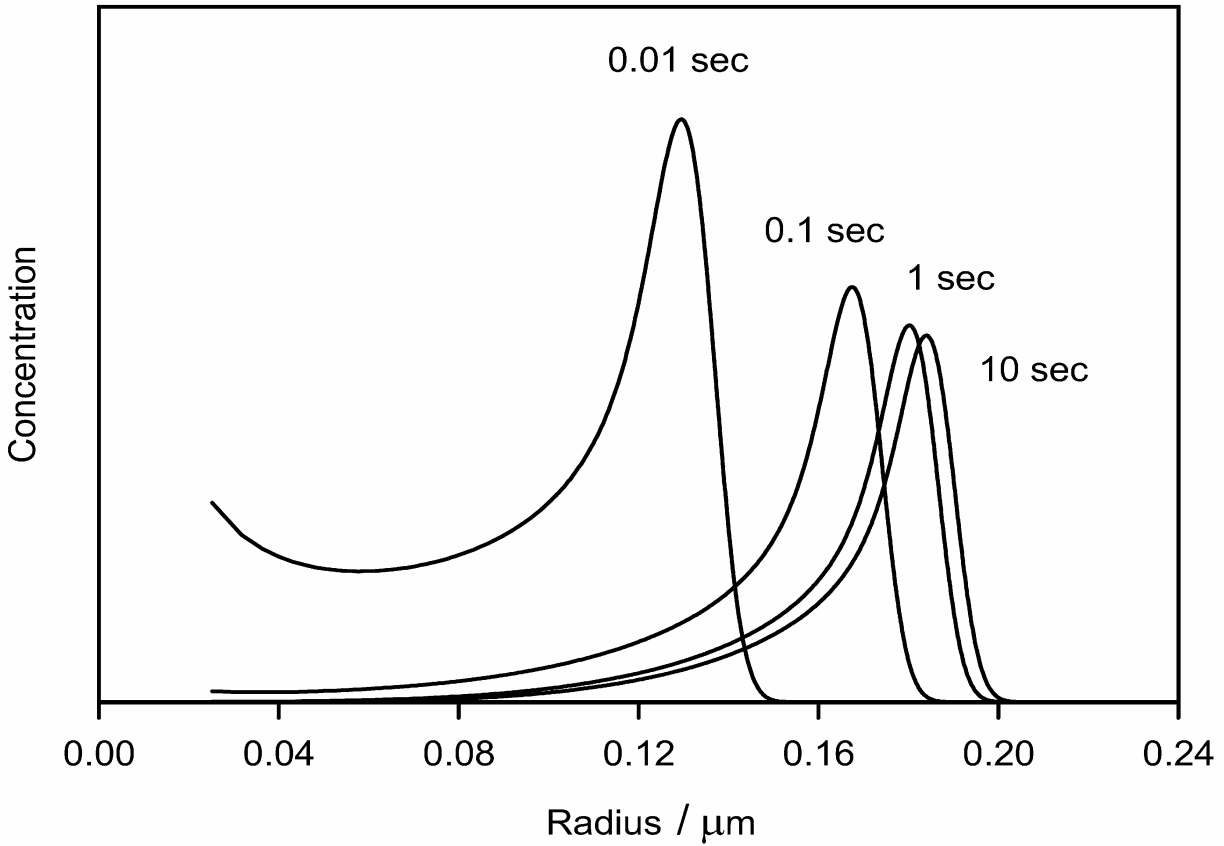

**Figure 4.** Example of a calculated colloid particle size distribution (in arbitrary units), plotted as a function of the colloid particle radius. The parameters correspond to a model of formation of spherical Au colloid particles, referenced in the text.

Another assumption in Eqs. (42-43) is that the radius of $s$-singlet, dense particles can be estimated as $R_p s^{1/3}$. However, primary particles actually have a distribution of radii, and they can also age (grow/coarsen) before their capture by and incorporation into the structure of the secondary particles. In order to partially compensate for the approximations, the following arguments are used. Regarding the size distribution of the singlets, it has been argued that since their capture rate especially by the larger aggregates is proportional to their radius times their diffusion constant, this rate will not be that sensitive to the particle size and size distribution, because the diffusion constant for each particle is inversely proportional to its radius. Thus, the product is well approximated by a single typical value.

The simplification of ignoring the primary particle ageing, was then further circumvented by using the experimentally determined typical primary particle linear size ("diameter"), $2R_{\exp}$, instead of attempting to estimate it as a function of time during the two-stage process. In fact, for the radius of the $s$-singlet particle, the expression in the first square brackets in Eq. (43), which represents the sum of such terms, $R_p s^{1/3}$, was recalculated with the replacement

$$R_p s^{1/3} \to 1.2 R_{\exp} s^{1/3}. \qquad (44)$$

Here the added factor is $(0.58)^{-1/3} \simeq 1.2$, where 0.58 is the filling factor of a random loose packing of spheres [58]. It was introduced to approximately account for that as the growing secondary particle compactifies by internal restructuring, not all its volume will be crystalline: A fraction will consists of amorphous "bridging regions" between the nanocrystalline subunits.

A possible inaccuracy in Eq. (39) because primary particles (those not yet captured) further grow by consuming additional solute matter, which, in fact, can be also directly consumed by the secondary particle surfaces, was partly compensated for [1] by renormalizing the distribution. This effect seems not to play a significant role in the dynamics. Some additional technical issues and details of the modeling are not reviewed here; see [1,3,5,7,8].

Two-stage models of the type just outlined, with singlet capture as the main growth mode of the secondary particles, were shown to provide a good semi-quantitative description (without adjustable parameters) of the processes of formation of spherical colloid-size particles of metals, Au [1,3,5,7,57,59] and Ag [57], a salt, CdS [7,8], as well as argued [60] to qualitatively explain the synthesis of an organic colloid — monodispersed microspheres of Insulin.

## 6. Further developments, and open problems

To improve the agreement between the results of the two-stage model and experimental data for secondary particle size distribution from semi-quantitative to quantitative, additional considerations were required. Here we begin by summarizing these developments, culminating in successful data fits for size distributions of CdS [7,8] and Au [59] particles, the former measured for different times during the process and for several protocols of feeding the solutes into the system, rather than just their instantaneous "batch" supply, as for the case illustrated in Figure 4.

Note that for non-batch supply of atomic-size "monomers," one has to add to the model the rate equations for their production in chemical reactions, which is, in itself, an interesting problem: Such reactions, involving the identification and modeling of the kinetics of various possible intermediate solute species, are not always well studied or understood theoretically, and they are not easy to probe experimentally.

In our numerical simulations, the parameters of the primary nucleation process, notably the value of the effective surface tension, were found to mostly affect the time scales of the secondary particle formation, i.e., the onset of "freezing" of their growth as illustrated in Figure 4. Accumulated evidence suggests that the use of the bulk surface tension and other experimentally determined parameters yields reasonable results consistent with the experimentally observed times.

The kinetic parameters of the secondary process seem to control primarily the average size of the final products. We found [1,3,5,7,8,59] that the particle sizes numerically calculated within the "minimal" model, while of the correct order of magnitude, were smaller than the experimentally observed values, by a non-negligible factor. The problem was traced to that the kinetics of the secondary aggregation, as described in Section 5, results in too many secondary particles which, since the total supply of matter is fixed, then grow to sizes smaller than those experimentally observed.

Two explanations for this effect were attempted. The first argued that for very small "secondary" aggregates, those consisting of one or few primary particles, the spherical-particle diffusional expressions for the rates, which are anyway somewhat ambiguous as described in connection with Eqs. (42-44), should be modified. Since the idea is to avoid introduction of many adjustable parameters, the rate $\Gamma_{1,1\to 2}$, cf. Eq. (43), was modified by a "bottleneck" factor, $f < 1$, with the underlying assumption that "merging" of two singlets (and other very small aggregates) may require substantial restructuring, thus reducing the rate of successful formation of a bi-crystalline entity. The two nanocrystals may instead unbind and diffuse apart, or merge into a single larger nanocrystal, effectively contributing to a new process, $\Gamma_{1,1\to 1}$, not in the original model. However, data fits [5,7,57] yield values of order $10^{-3}$ or smaller for $f$, which seems too drastic a reduction factor.

Another modification of the model uses a similar line of argument but in a somewhat different context. We point out that the model already assumes a certain "bottleneck" for particle merger — that of singlet-capture dominance. Indeed, all the rates in Eq. (43) with both $s_1 > 1$ and $s_2 > 1$, are set to zero. This assumption was made based on empirical experimental observations that larger particles were never seen to pair-wise "merge" in solution. It seems that the restructuring process that causes rapid compactification of the growing secondary particles, and which is presently not understood experimentally or theoretically, can also cause incorporation of primary particles, but not larger aggregates, in the evolving structure, while retaining their crystalline core to yield the final polycrystalline colloids.

One might then argue that perhaps small aggregates, up to certain cutoff sizes, $s_{\max} > 1$, can also be dynamically rapidly incorporated into larger aggregates on diffusional encounters. Thus, we can generalize the model equations, see [7,8] for details, to allow for cluster-cluster aggregation with rates given by Eq. (43), but only as long as at least one of the cluster sizes does not exceed a certain value $s_{\max}$. The sharp cutoff is an approximation, but it offers the convenience of a single new adjustable parameter. Indeed, data fits for CdS and Au spherical particles, yield good quantitative agreement, exemplified in Figure 5, with values of $s_{\max}$ ranging from 15 for Au, to 25 for CdS. Interestingly, these values are not only intuitively reasonable as defining "small" aggregates, but they also fit well with the concept of the cutoff value $n_{\text{th}}$, discussed in Section 4, only beyond which atomistic aggregates develop a well formed "bulk-like" core. Indeed, the only available numerical estimate of such a quantity in solution [37], for AgBr nano-aggregates, suggests that $n_{\text{th}}$ is comparable to or somewhat larger than 18 (in terms of molecule count, i.e., the most stable configuration for a $Ag_{18}Br_{18}$ nanocluster is disordered). We also comment that cluster-cluster aggregation at small sizes, can explain the formation of the initial peak in the secondary-particle distribution, which later grows by the fast-capture-of-singlets mechanism.

The modification/elaboration of the two-stage model just outlined, required large-scale numerical effort and lead to development of adaptive-mesh (in time and cluster size) algorithmic techniques for efficient simulations [7,8].

Finally, we point out that in the described treatments we avoided any quantitative or even qualitative modeling of the particle (nanosized and colloid) shape selection. Many of the processes that could be treated in a cavalier way in studying the particle sizes will balance to determine the details of the shape distribution of the final products. These processes include particle restructuring, both in the interior and at surfaces, as well as monomer transport on particle surfaces and possible monomer detachment/reattachment, as well as detachment/attachment/reattachment/surface motion of larger than monomer structures. The difficulty in modeling these processes is two-fold. Firstly, they are presently not quantified and are difficult to probe experimentally. Secondly, their modeling would require extremely large-scale simulations. Thus, while one can venture guesses as to the key processes that balance to determine the particle shape distribution, derivation of quantitative predictions and their comparison with experimental data remain an important open challenge in colloid and nanoparticles science.

The author acknowledges funding of this research by the U.S. National Science Foundation (grant DMR-0509104).

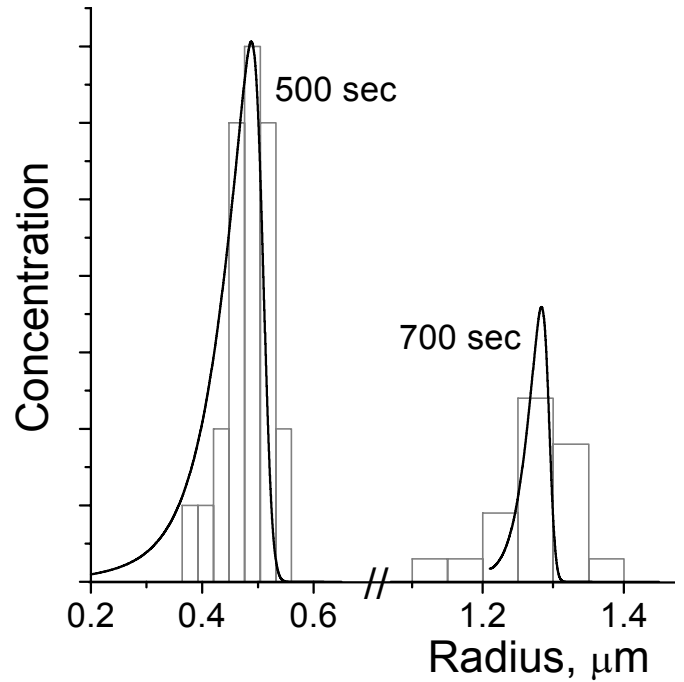

**Figure 5.** The calculated (curves) and experimentally measured (histograms) particle size distributions (in arbitrary units), for two different times during the growth, plotted as functions of the particle radius. The parameters correspond to the $s_{\max} = 25$ model of formation of spherical CdS colloids.